\documentclass{ar-1col-S2O}
\usepackage[numbers]{natbib}
\usepackage{amsmath}
\usepackage{amsfonts}
\usepackage{url}
\usepackage{adjustbox}
\def\be{\begin{equation}} 
\def\ee{\end{equation}}
\newcommand \bea {\begin{eqnarray}} 
\newcommand \eea {\end{eqnarray}}

\jname{Xxxx. Xxx. Xxx. Xxx.}
\jvol{AA}
\jyear{YYYY}
\doi{10.1146/((please add article doi))}

\begin{document}

% Page header
\markboth{Yampolskaya and Mehta}{Hopfield Models}

% Title
\title{Hopfield Networks as Models of Emergent Function in Biology}
%Authors, affiliations address.
\author{Maria Yampolskaya,$^{1*}$ Pankaj Mehta$^{1,2,3,4,\dagger}$
\affil{$^1$Department of Physics, Boston University, Boston, MA, USA, 02215}
\affil{$^2$Center for Regenerative Medicine of Boston University
and Boston Medical Center, Boston, MA, USA 02215}
\affil{$^3$Faculty of Computing and Data Science, Boston University, Boston, MA, USA 02215}
\affil{$^4$Biological Design Center, Boston University, Boston, MA, USA 02215}
\affil{$^*$email: mariay@bu.edu}
\affil{$^{\dagger}$email: pankajm@bu.edu}
}

%Abstract
\begin{abstract}
Hopfield models, originally developed to study memory retrieval in neural networks, have become versatile tools for modeling diverse biological systems in which function emerges from collective dynamics. In this review, we provide a pedagogical introduction to both classical and modern Hopfield networks from a biophysical perspective. After presenting the underlying mathematics, we build physical intuition through three complementary interpretations of Hopfield dynamics: as noise discrimination, as a geometric construction defining a natural coordinate system in pattern space, and as gradient-like descent on an energy landscape. We then survey recent applications of Hopfield networks a variety of biological setting including cellular differentiation and epigenetic memory, molecular self-assembly, and spatial neural representations.
\end{abstract}

%Keywords, etc.
\maketitle

%Table of Contents
\tableofcontents

\section{Introduction}
From the millions of proteins in a single cell to the billions of neurons in the human brain, collective behavior is responsible for much of the complex phenomena observed in biology \cite{haddow1945life,alberts1998cell,nelson2004biological,phillips2006biological,phillips2012physical,bialek2012biophysics}. Across scales, biological systems perform specific functions as a result of many parts working together. Modeling the emergence of function from these components is part of what makes biophysics so difficult \cite{hopfield2014whatever}.  One promising approach for meeting this challenge is to recast this problem in the language of high-dimensional dynamical systems. The most famous and fruitful example of this approach is the Hopfield model \cite{hopfield1982neural}.  Originally conceived as a model of memory in neural networks, the Hopfield model has now found broad applicability across biophysics, including understanding the epigenetics of cellular identity and differentiation, modeling the self-assembly of heterogeneous structures,  and thinking about neural representations of spatial environments in the hippocampus.

The Hopfield model has a long pedigree, and many papers have applied analytical tools from statistical physics to calculate properties of these models, including phase diagrams and storage capacities \cite{amit1985storing,PhysRevA.32.1007, amit1989modeling, mezard2017mean, hopfield1982neural,krotov2016dense,demircigil2017model,https://doi.org/10.48550/arxiv.2304.14964,Agliari2023}. Recently, Hopfield models have attracted renewed interest due to their close relationship with modern machine learning techniques such as transformers \cite{ramsauer2020hopfield} (see \cite{krotov2023new} for a recent review that highlights these developments). Here, we review the Hopfield networks from a biophysical perspective, focusing on Hopfield models as exemplars of complex systems exhibiting emergent function \cite{amit1989modeling,hopfield2010neurodynamics,krotov2016dense,demircigil2017model,zhong2017associative,krotov2020large,ramsauer2020hopfield,karin2024enhancernet}. We explain the intuition behind Hopfield dynamics, why and how these models work, and how they can be used to understand disparate biophysical phenomena. 

The review is organized as follows. In Section~\ref{mathematical}, we present mathematical descriptions of classic and modern Hopfield networks. In Section~\ref{interpretations}, we provide three biophysically inspired perspectives for interpreting the dynamics of Hopfield models: (i) as a system that discriminates between signal and noise; (ii) as a geometric operation that projects onto a subspace; and (iii) as gradient-like dynamics on a landscape.  We also include Section~\ref{other}, which provides an overview of other, more commonly, approaches for thinking about Hopfield models. In Section~\ref{biophysics}, we turn to applications and describe how Hopfield models have been used to study a variety of biophysical phenomena including cell fate transitions, molecular self-assembly, and neural representations.

\section{The Mathematics of  Hopfield models}\label{mathematical}
 
  \subsection{Overview}

 In this section, we provide a brief self-contained introduction to the mathematics of Hopfield models. We start by introducing the basic mathematical objects used to describe Hopfield models. We then specialize our discussion to three commonly used variants of the Hopfield model: the classic network introduced by Hopfield  \cite{hopfield1982neural}, the projection method \cite{Personnaz1985, Kanter1987}, and the exponential ``Modern Hopfield Network'' \cite{krotov2016dense,demircigil2017model,ramsauer2020hopfield}. Despite the many implementational differences between traditional and modern networks, they share a common underlying logic, namely: (i) a dynamical update for the state of the system (ii) and a set of stored patterns that act as attractors (fixed points) of the dynamics. When the dynamics are initialized sufficiently close to a stored pattern, the system flows to the corresponding fixed point, allowing the network to ``retrieve'' the stored pattern.  

\begin{figure}[t]
\includegraphics[width=\textwidth]{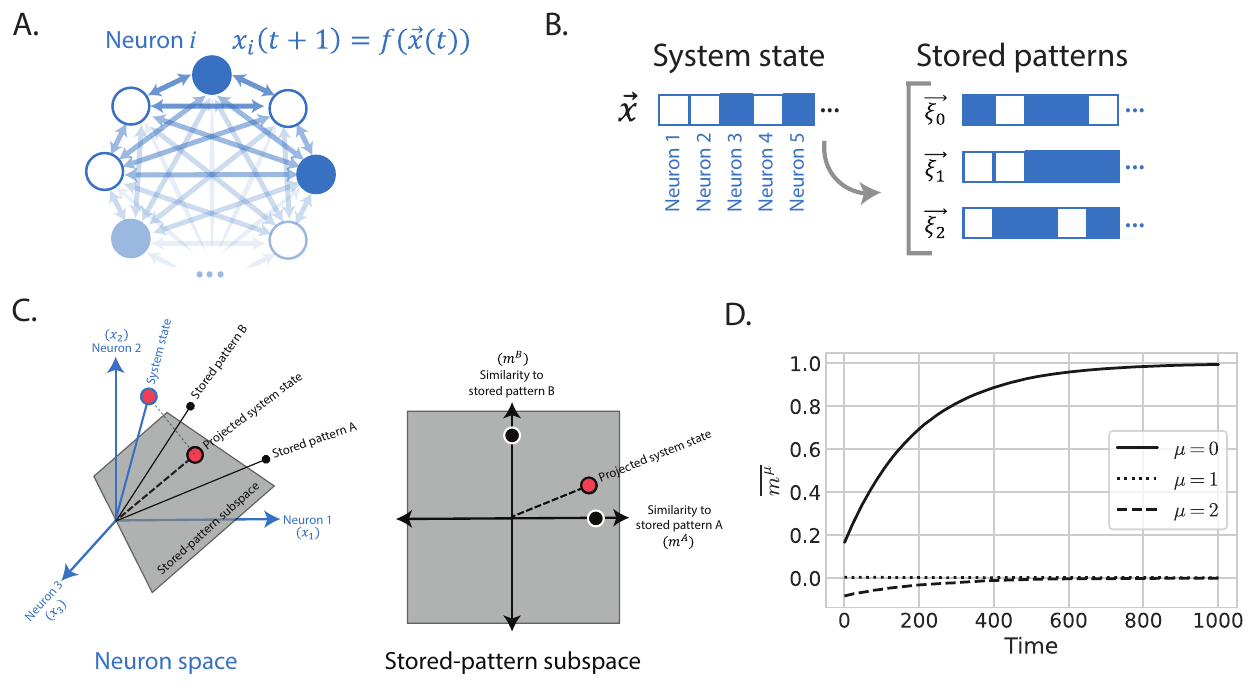}
\caption{Interpretations of Hopfield models. A. Hopfield models involve networks of neurons, and the update rule for each neuron depends on the state of every other neuron. In classic Hopfield networks, each neuron has a binary value, and neurons are connected all-to-all with pairwise interactions. B. Hopfield models are also called associative memory models because they compare the system state, $\vec{x}$, to a list of memories, or stored patterns, and retrieve the memory that has the most association with the initial state. States are vectors in neuron space. In this illustration, the neurons take on binary values, though Hopfield models are flexible enough to allow continuous values as well. C. The generalized order parameters $\{m^{\mu}\}$ in Equation \ref{eqn:generalized order param} can be interpreted as coordinates defining stored-pattern space. They are the result of projecting from the higher-dimensional neuron space down to the subspace spanned by stored patterns. [Figure adapted from \citet{yampolskaya2023sctop} [Note to Annual Reviews: we are authors of this article; the publisher grants authors the right to reuse their own figures without permission.]] D. The dynamics of the generalized order parameters for a classic Hopfield network with 3 patterns, 200 neurons, and low temperature ($T = 0.01$). The order parameters are averaged over 500 iterations of the simulation. Each simulation is started with the same initial state, which has some overlap with the $0$th stored pattern, $\xi_{0}$. As the simulation progresses, the network retrieves the $0$th stored pattern, so $m^0$ becomes 1 (indicating perfect alignment) and $m^{\mu \neq 0}$ become $0$ (indicating no alignment with those stored states). }
\label{fig:interpretations}
\end{figure}

Hopfield models share a set of common ingredients (see Fig. \ref{fig:interpretations}):
 
 \begin{itemize}
    \item A $N$-dimensional state vector $\vec{x}(t)=(x_1(t),\ldots,x_N(t))$ whose $i$-th component $x_i(t)$ encodes the state of the variable $i$-th dynamical variable at a time $t$.  The $x_i$  can be binary-valued (e.g., $x_i = \pm 1$) or continuous (e.g., $x_i\in \mathbb{R}$). In the computational neuroscience and physics literatures, these variables are often called ``neurons" or ``spins''.     
        
 \item A  matrix $\xi_{\mu i}$ with dimensions $P \times N$  describing the $P$ patterns we wish to store.  By construction, when the system is in a fixed point corresponding to the $\nu$-th stored pattern, $x_i = \xi_{\nu i}$. In the spin glass literature, these stored patterns are often assumed to be random and independent. However, in many biophysical applications we discuss below, it will be important to consider the case where stored patterns are correlated.
  
 \item An update rule or differential equation governing the dynamics, e.g. $x_i(t+1) = f_i(\vec{x}(t))$ for discrete time or $\frac{dx_i}{dt} = f_i(\vec{x}(t)) - x_i$ for continuous time. In the Hopfield model, the functions $f_i(\vec{x}(t))$ are chosen in such a way as to ensure that all $P$ patterns $\{ \vec{\xi}_\nu\}$ are fixed points of the dynamics when the number of patterns $P$ is sufficiently small compared to the number of neurons $N$. 
  
 \item A $P$-dimensional ``order parameter'' vector $\mathbf{m}(t)=(m_1(t), \ldots, m_P(t))$ that measures the alignment of the state of the system $\vec{x}(t)$ at a time $t$ with each of the $P$ stored patterns. As discussed below, describing the systems in terms of order parameter offers a rich lens for interpreting Hopfield dynamics.
 
 \item An energy function, or Lyapunov function, which decreases with time as a result of the update rule.
\end{itemize}

What distinguishes model variants from each other is how these basic ingredients are combined to construct the detailed network dynamics. Some models consider discrete variables where $x_i=\pm 1$, whereas others use continuous $x_i$. Another major difference between model variants is how the stored patterns $\xi_{\mu i}$ are constructed. The classic Hopfield model assumes these are binary random vectors whereas other constructions relax this assumption. The biggest difference between model variants is the choice of non-linear function $f_i(\vec{x}(t))$ that defines the update rule. Here, we focus on three choices for this function corresponding to the classical Hopfield model, the projection method, and Modern Hopfield networks.

\subsection{Classic Hopfield Network with random patterns}\label{classicHopfield}

In 1982, Hopfield formulated his now classic network as a model of collective computation \cite{hopfield1982neural, doi:10.1073/pnas.81.10.3088}. The network consists of $N$ binary-valued neurons that interact with each other through a pairwise interaction matrix $J_{ij}$ (see Figure \ref{fig:interpretations}). At a time t, a neuron can be firing ($x_i(t)=1$) or silent ($x_i(t)=-1$). At each time step, a randomly chosen neuron is updated using an asynchronous dynamical update rule of the form
\begin{equation}
  x_i (t+1) = \text{sign} (\sum_{j\neq i} J_{ij} x_j(t))
\label{Eq:Hop-dyn-gen}
\end{equation}
The key insight of Hopfield was the realization that it is possible to choose the couplings $J_{ij}$ in such a way as to ensure that a specified set of binary patterns (memories) $\{\vec{\xi}_{\mu} \}$ are fixed points of the dynamics (see Figure \ref{fig:interpretations}). For the special case of random  $\xi_{\mu i}$, Hopfield showed that choosing the couplings to be of the form
\begin{equation}
J_{ij} = \frac{1}{N} \sum_{\mu=1}^P \xi_{\mu i} \xi_{\mu j}, 
\label{Eq:Hop-J-gen}
\end{equation}
ensures that the the patterns $\{\vec{\xi}_{\mu} \}$ are fixed points of the dynamics as long as the number of patterns is much smaller than the number of neurons, $P \ll N$. We reproduce this argument  in Sec. \ref{signal}.  

Hopfield also showed that these dynamics possess a Lyapunov function,
\begin{equation}
 E = - \frac{1}{2} \sum _{ij}x_i J_{ij} x_j,
 \label{Eq:Hop-E-gen}
\end{equation}
To see that $E$ monotonically decreases under the  update rule in Eq.~\ref{Eq:Hop-dyn-gen}, consider the energy difference between two states that differ at the $k$th spin:
\begin{align*}
    \Delta E &= E(x_k= +1) - E(x_k=-1) \\
    &= (-\frac{1}{2} \sum_{i\neq k, j\neq k} x_i J_{ij} x_j -  \sum_{j} (+1) J_{kj} x_j)  - (-\frac{1}{2} \sum_{i\neq k, j\neq k} x_i J_{ij} x_j -  \sum_{j} (-1) J_{kj} x_j ) \\
    &= - 2\sum_{j}J_{kj} x_j
\end{align*}
If the state with $x_k=+1$ has lower energy, $\Delta E = -\sum_{j}J_{kj} x_j < 0$ and $\text{sign}(\sum_{j}J_{kj} x_j) > 0$, so the update rule returns $x_k (t+1) = +1$. If the state with $x_k=-1$ has lower energy, $\Delta E = -\sum_{j}J_{kj} x_j > 0$ and $\text{sign}(\sum_{j}J_{kj} x_j) < 0$, so the update rule returns $x_k (t+1) = -1$.   

Hopfield's paper set off a flurry of activity among statistical physicists (see \cite{amit1989modeling} for a pedagogical and comprehensive account). One powerful concept that emerged from this spin-glass-inspired perspectives was a description of these systems in terms of order parameters $m_{\mu}$
that measure the overlap between the current state of the system and stored patterns. For the classic Hopfield network, these order parameters take the form of generalized magnetizations,
\begin{equation}
m_{\mu} = \frac{1}{N} \sum_{j} \xi_{\mu j} x_j.
\label{Eq:Hop-def-m}
\end{equation}
Both the dynamical update rule and the Lyapunov function $E$ can be re-expressed in terms of order parameters. Using  Eqs.~\ref{Eq:Hop-J-gen}, it is easy to show that the energy function takes the simple form
\begin{align*}
    E = -\frac{N}{2} \sum_{\mu} (m_{\mu})^2
\label{Eq:Hop-E-m}
  \end{align*}
We will make use of this expression for $E$ to interpret the dynamics of the Hopfield model in terms of energy landscapes in Section~\ref{landscape}. 

The dynamic update rule (Eq.~\ref{Eq:Hop-dyn-gen}) can be rewritten in terms of $m_\mu$ as
\begin{equation}
    x_i (t+1)= \text{sign} (\sum_{\mu} \xi_{\mu i} m_{\mu} (t)) 
 \label{Eq:Hop-dyn-m}
\end{equation}
This form of the dynamical update rule makes it clear that the state of the system is pushed towards  stored patterns with which it has higher overlaps (i.e. patterns with large $m_\mu$). The nonlinearity in the update rule-- in this case, the $\text{sign}$ function -- forces the system to ``choose'' between patterns.  However, the almost linear nature of this update rule often results in the presence of spurious attractors -- fixed points of the dynamics that are not the stored patterns. The most common spurious attractors are odd mixtures of existing patterns of the form
$\vec{\xi}_{mix}= \pm \text{sign}(\pm \vec{\xi}_{\mu_1}\pm \vec{\xi}_{\mu_2}\pm \vec{\xi}_{\mu_2})$. Alternative learning rules with different non-linearities can be constructed that  further reduce the effect of spurious attractors by decreasing the energy of the system in those states \cite{Hopfield1983,VanHemmen1990,Serricchio2025}.

Spurious attractors play an important role in the study of Hopfield networks. As the number of stored patterns $P$ increases, the form of the dynamical update rule Eq.~\ref{Eq:Hop-dyn-m} suggests the system will be pulled in  more and more distinct directions because the state of the system overlaps with multiple stored patterns. This frustration in the dynamics ultimately places a limit on the number of stored patterns that can be successfully recalled, giving rise to the concept of a storage capacity. In classical Hopfield models, the storage capacity $P_{max}$ scales \emph{linearly} with the number of neurons $N$ \cite{hopfield1982neural, amit1985storing, amit1989modeling}. We provide an intuitive explanation for this result in Section~\ref{signal}.

Finally, thus far we have limited our discussion to discrete-time updates. However, it is straightforward to formulate the dynamics of Hopfield networks in continuous time. To do so, one simply approximates the discrete update rule using the derivatives as follows:
\begin{align*}
    \frac{dx_i}{dt} &\approx \frac{x_i (t + \Delta t) - x_i (t)}{ \Delta t} \\
    &\approx \frac{1}{\tau} (x_i(t+1) - x_i (t))\\
    &= \frac{1}{\tau} (\text{sign} (\sum_{j\neq i} J_{ij} x_j(t))-x_i(t)),
\end{align*}
where $x_i(t+1)$ is the discrete-time update rule, $\tau$ is a constant that sets the time scale for the dynamics, and in going from the second to third equation we have used Eq.~\ref{Eq:Hop-dyn-gen}.

\subsection{Storing correlated patterns with the projection method}
Up to this point, we have considered a classic Hopfield model with random patterns. However, when the patterns are correlated, the classic Hopfield network works poorly. The reason for this is that in this case,  the system state $\vec{x}$ will generically have large overlaps with multiple patterns. Namely, if two patterns $\vec{\xi}_\mu$ and $\vec{\xi}_{\nu}$ are highly correlated, then Eq.~\ref{Eq:Hop-def-m} implies that if $m_\mu$ is large, then $m_\nu$ must also be large. This interference between patterns gives rise to frustration in the dynamics because the system is pulled in multiple directions by the dynamic update rule Eq.~\ref{Eq:Hop-dyn-m}. This results in increased errors in pattern retrieval and low storage capacity.

\citet{Kanter1987} proposed a simple modification of the classic Hopfield scheme, the projection method,  that allows for pattern retrieval even when patterns are correlated. The key insight of this work was to modify Eq.~\ref{Eq:Hop-J-gen}, the rule for constructing neural couplings $J_{ij}$, to account for correlations between patterns. In the projection method, one starts by defining the correlation matrix between patterns
\begin{equation}
g_{\mu \nu} = \sum_{ij} \xi_{\mu i} \xi_{\nu j},
\label{Eq:g-def}
\end{equation}
as well as the inverse of this matrix
\begin{equation}
g^{\mu \nu} = g_{\mu \nu}^{-1}.
\end{equation}
The dynamics are still governed by Eq.~\ref{Eq:Hop-dyn-gen}, but now with the couplings defined according to the 
``projection rule''
\begin{equation}
J_{ij} = \frac{1}{N} \sum_{\mu,\nu=1}^P \xi_{\mu i} g^{\mu \nu} \xi_{\nu j}.
\label{Eq:Hop-J-projection}
\end{equation}
We give an intuitive explanation of the origins of this rule and why it works in Section~\ref{projection}.

It is also helpful to define ``decorrelated'' order parameters
\begin{equation}
    m^{\mu} = \sum_{\nu} g^{\mu \nu} m_{\nu},     
\label{Eq:def-m-up}
\end{equation}
that generalize the usual magnetizations (Eq.~\ref{Eq:Hop-def-m}). Note that we use a raised index to distinguish this decorrelated order parameter from magnetizations. When the stored patterns are orthogonal (i.e. $g_{\mu \nu} = \delta_{\mu \nu}$), the generalized order parameters are identical to the classic one ($m^{\mu} = m_{\mu}$). A straight forward calculation shows that for coupling as in Eq.~\ref{Eq:Hop-J-projection}, the Lyapunov function (Eq.~\ref{Eq:Hop-E-gen}) takes the simple form \cite{Personnaz1985,Kanter1987}:
\begin{eqnarray}
E &=& -\frac{N}{2} \sum_{\mu} m^{\mu} g^{\mu \nu} m_{\nu} \nonumber \\
E &=& -\frac{N}{2} \sum_{\mu} m^{\mu} m_{\mu}. 
\label{Eq:Proj-E-m}
\end{eqnarray}
Written in this form, it's clear that the energy is a dot product of the states in order parameter space, with the metric tensor correcting for the distortion of angles due to the non-orthogonality of patterns. Unless otherwise stated, we assume that patterns are correlated. For this reason, we will make extensive use of both generalized order parameters and magnetizations in what follows.

\subsection{Modern Hopfield Networks}

The final class of models we will discuss are a large class of networks that are commonly called  ``modern Hopfield networks''.  Modern Hopfield networks generalize classic Hopfield models by introducing higher-order interactions between spins  \cite{Gardner1987,krotov2016dense,krotov2020large}. Unlike classic Hopfield models where the energy function is second-order in the spins (see Eq~\ref{Eq:Hop-E-gen}), in modern Hopfield networks the energy function can have terms of any order.

In this review, we restrict our discussion to exponential Modern Hopfield  networks where  the dynamic update rule and energy function take the form \cite{demircigil2017model,ramsauer2020hopfield,https://doi.org/10.48550/arxiv.2304.14964}. 
\begin{eqnarray}
    x_i (t+1) &=& \sum_{\mu} \xi_{\mu i} \sigma^{\mu}(\beta m^{\mu}) \nonumber \\
    E &=& -\sum \log (\sum_{\mu} \exp(m^{\mu})) + \frac{1}{2} \sum_{i}x_i^2,
\label{eqn:softmax energy}
\end{eqnarray}
with $\sigma$ the soft-max function with an inverse temperature parameter $\beta$,
\begin{equation}
\sigma^{\mu}(\beta m_{\mu}) = \frac{e^{\beta m_{\mu}}}{\sum_{\nu} e^{\beta m_{\nu}}},
\end{equation}
and $m^\mu$ is as in  Eq.~\ref{Eq:def-m-up}. A visualization of the energy function is shown in Figure~\ref{fig:softmax energy}A.  One of the advantages of these dynamics is that unlike the classic Hopfield network, both the neurons $\vec{x}$ and the patterns $\vec{\xi}$ can be real-valued.

For sufficiently small temperatures, it is easy to show that the stored patterns $\xi_{\mu i}$ are fixed points of the dynamics. To see this, it is helpful to rewrite the dynamic update rule entirely in terms of generalized order parameters. To do this, we multiply both side of the dynamic update rule by $\sum_\nu g^{\gamma \nu} \xi_{\nu i}$ and sum over $i$ to get:
\begin{eqnarray}
\sum_{i,\nu} g^{\gamma \nu} \xi_{\nu i}  x_i (t+1) &=& \sum_{\mu,\nu} g^{\gamma \nu} \sum_i \xi_{\nu i} \xi_{\mu i} \sigma^{\mu}(\beta m^{\mu}) \nonumber \\
\sum_{\nu} g^{\gamma \nu} m_\nu(t+1) &=& \sum_{\mu,\nu} g^{\gamma \nu} g_{\nu \mu} \sigma^{\mu}(\beta m^{\mu})\nonumber \\
m^{\mu}&=&\sigma^{\mu}(\beta m^{\mu}),
\label{Eq:m-modern-dynamics}
\end{eqnarray}
where, in going to second line, we have used Eqs.~\ref{Eq:Hop-def-m} and ~\ref{Eq:g-def}, and in the third line we have used Eq.~\ref{Eq:def-m-up} and the fact that $g^{\gamma \nu}$ is the inverse of $g_{\nu \mu}$. Notice that in the zero temperature limit $\beta \rightarrow 0$, any order-parameter vector $\mathbf{m}=(0, \ldots,0, 1,0, \ldots,0)$ with all $m_{\mu}=0$ except for a single pattern $\gamma$ is a fixed point of the dynamics because  $\sigma^{\mu}(\beta m^{\mu}) \approx 0$ in $\mu \neq \gamma$ and $\sigma^{\mu}(\beta m^{\mu}) \approx 1$ if $\nu =\gamma$. Thus just like in the classic Hopfield networks, the network retrieves the stored patterns it is closest to. 

The update rule for modern Hopfield networks bears some striking similarities to the classic Hopfield construction. It involves a nonlinear function moving the system towards the state with the highest alignment $m^{\mu}$. Now, however, the nonlinearity is a soft-max instead of a sign function, leading to exponential storage capacity \cite{https://doi.org/10.48550/arxiv.2304.14964}. Additionally, the stored patterns are outside of the nonlinearity in the sum, which allows for continuous-valued stored patterns. Table \ref{table:classic_modern} provides a comparison of modern and classical networks.

\begin{table}[t!]
\tabcolsep5pt
\label{table:classic_modern}
\begin{center}
\adjustbox{width=\textwidth}{
\begin{tabular}{|l|c|c|}\hline
\small
 &Classic&Exponential Modern\\
\hline
 State variables&Typically binary ($x_i = \pm 1$)&Typically continuous ($x_i \in \mathbb{R}^N$)\\\hline
 Dynamics&$x_i (t+1) = \text{sign} (\sum_{\mu} \xi_{\mu i} m_{\mu}(t) $& $x_i (t+1) = \sum_{\mu} \xi_{\mu i} \sigma^{\mu}(\beta m^{\mu} (t))$\\\hline
 Energy function&$E = - \frac{1}{2} \sum_{ij}x_i J_{ij} x_j  = - \frac{1}{2} \sum_{\mu} (m_{\mu})^2$&$ E = -\sum \log (\sum_{\mu} \exp(m^{\mu})) + \frac{1}{2} \sum_{i}x_i^2 $\\\hline
 Pattern storage capacity&Linear in the number of neurons ($P_{\max} \propto N$)&Exponential in the number of neurons ($P_{\max} \propto 2^N$)\\
\hline
\end{tabular}}
\adjustbox{max width=\columnwidth}{...}
\caption{A comparison of classical and modern Hopfield networks.}
\end{center}

% \begin{tabnote}
% $^{\rm a}$Table footnote; $^{\rm b}$second table footnote.
% \end{tabnote}
\end{table}

\section{Interpretations of Hopfield dynamics}\label{interpretations}

The previous section introduced the mathematics of the Hopfield model. In this section, we focus on developing intuition. To do so, we present different perspectives through which one can view these systems, each of which emphasizes a different aspect of these networks.

\subsection{As retrieving a signal}\label{signal}
One way of interpreting Hopfield dynamics is as a system that discriminates between a ``target" pattern -- the pattern it is most aligned with, and is expected to converge to -- and the ``non-target" patterns. In other words, the system picks out the signal of the intended pattern from the noise caused by interference from other patterns. This interference is due to random overlaps between stored patterns, which become more significant when there are a large number of patterns relative to the number of neurons. For simplicity, we will demonstrate this concept using the classic Hopfield model with orthogonal stored patterns, but the intuition presented here holds in more general settings. 

Suppose the system is in the stored state corresponding to pattern $\nu$ ($x_i = \xi_{\nu i}$). The correct behavior of the system is to remain in the target pattern $\nu$ and not to flow to other patterns. However, if the number of stored patterns $P$ is close to the number of neurons $N$, interference from non-target patterns can cause errors in retrieval, and even destabilize the fixed point corresponding to the target pattern. To see this, consider the update rule in Eq.~\ref{Eq:Hop-dyn-gen} with couplings given by Eq.~\ref{Eq:Hop-J-gen}. When the system state is aligned with the $\nu$-th pattern, $\vec{x}(t)=\vec{\xi}_{\nu}$, the update rule takes the form:
  \begin{align*}
        x_i (t+1) &= \text{sign} (\frac{1}{N} \sum_{\mu, j} \xi_{\mu i} \xi_{\mu j}  \xi_{\nu j}) \\
        &= \text{sign} (\frac{1}{N} ( \sum_j \xi_{\nu i} \xi_{\nu j}  \xi_{\nu j} + \sum_{\mu \neq \nu, j} \xi_{\mu i} \xi_{\mu j}  \xi_{\nu j})) \\
        &= \text{sign} (\frac{1}{N} (N \xi_{\nu i}  + \sum_{\mu \neq \nu, j} \xi_{\mu i} \xi_{\mu j}  \xi_{\nu j})) \\
        &= \text{sign} (\xi_{\nu i}  + \frac{1}{N}\sum_{\mu \neq \nu, j} \xi_{\mu i} \xi_{\mu j}  \xi_{\nu j}).
  \end{align*}
The first term inside the sign function is just the target pattern and can be thought of as the ``signal''. The second term is an interference term that comes from the non-target patterns and can be thought  of as ``noise''. Whether or not the dynamics remain in the correct pattern depends on the relative magnitude of these two terms. When the second term becomes non-negligible, there will be errors in pattern storage and retrieval. In extreme cases, this noise term overwhelms the signal and the pattern is no longer stable.

Since $\xi_{\mu i}$ are drawn randomly, the interference term is a sum of the product of three random variables. Because $\xi_{\mu i} = \pm 1$, this sum is like a random walk over $(P-1)N$ random steps. The error caused by this interference term is the variance of a random walk (equivalently, it is the variance of a sum of random variables, so one can apply the central limit theorem). We can rewrite the sum in terms of the mean $\mu_w$ (which is zero for a random walk), variance $\sigma^w$, and a random normally-distributed variable $z \sim N(0,1)$:
\begin{align*}
    \sum_{\mu \neq \nu, j}^{P, N} \xi_{\mu i} \xi_{\mu j}  \xi_{\nu j} &= \mu_w + \sigma^w z\\
    &= 0 + z\sqrt{(P-1)N}.
\end{align*}
This yields the following equation for the update rule.
\begin{align*}
    x_i (t+1) &= \text{sign} (\xi_{\nu i}  + \frac{\sqrt{(P-1) N}}{N}z) \\
    &\approx \text{sign} (\xi_{\nu i}  + \sqrt{\frac{P}{N}}z)
\end{align*}
The error in memory retrieval scales like $\sqrt{\frac{P}{N}}$. For this reason, the number of network neurons must be much larger than the number of stored patterns to keep the errors low. This expression also shows that the storage capacity scales linearly with the number of neurons, $P_{\max} \propto N$. An analogous argument can be formulated for Modern Hopfield networks and one finds that the storage capacity is exponential in the number of neurons $N$ \cite{demircigil2017model,ramsauer2020hopfield,https://doi.org/10.48550/arxiv.2304.14964}.

\subsection{As projection onto a subspace}\label{projection}
Another insight into Hopfield models comes from recognizing that the transformation from the state space of the system $x_i$ to  order parameters $m^{\mu}$ can be viewed as change of basis from  ``neuron space'' to ``pattern space''. This way of thinking about Hopfield models is especially powerful when the number of patterns is less than the number of neurons $P\le N$. As shown in Figure \ref{fig:interpretations}B, in this case, the patterns span a subspace of the full neuron space and whose natural ``basis'' is the patterns themselves. 

In particular, we can construct a projection matrix $P$ that acts on a state $\vec{x}$ and projects it down to the subspace spanned by the stored patterns. From basic linear algebra, we know that this projection matrix takes the form
\begin{equation}
P_{ij}=\sum_{\mu, \nu} \xi_{\mu i} g^{\mu \nu}\xi_{\nu j},
\label{Eq:def-P}
\end{equation}
where $g_{\mu \nu}$ is given by Eq.~\ref{Eq:g-def}. It is easy to verify that this is indeed a projection matrix by confirming that $P^2=P$. In terms of the projection matrix, we can decompose a vector $\vec{x}$ into a component that lies in the pattern subspace and a component orthogonal to this subspace,
\be
\vec{x} = P \vec{x}+ \vec{x}^{\perp},
\ee
where by definition $\vec{x}^{\perp}=(I-P)\vec{x}$. Using Eq.~\ref{Eq:def-m-up}, this decomposition can be written as
\be
 x_i   =  \sum_{\mu} m^{\mu} \xi_{\mu i} + x^{\perp}_i.
\ee
This expression shows that the generalized order parameters $m^{\mu}$ are the natural coordinate system for describing the state of the system in terms of patterns. The projection method gets its name from the fact that the couplings $J_{ij}$ defined in Eq.~\ref{Eq:Hop-J-projection} are exactly the projection matrix $J_{ij}=P_{ij}$. With this in mind, we see that the energy of classic Hopfield model in Eq.~\ref{Eq:Proj-E-m} is simply the magnitude of the orthogonal component of the current state, namely $E=\vec{x}^{\perp}\cdot \vec{x}^{\perp}$.

Thus, far we focused our discussion on classic Hopfield models. However, this perspective also sheds light on dynamics of modern Hopfield networks. By substituting Eq.~\ref{Eq:m-modern-dynamics} into Eq.~\ref{eqn:softmax energy}, we can rewrite the dynamics of the modern Hopfield network as
\be
x_i (t+1) = \sum_{\mu} \xi_{\mu i} m^\mu(t+1).
\ee
This equation emphasizes one of the key intuitions behind modern Hopfield networks: the natural space for thinking about dynamics is not neuron space, $\vec{x}$, but  pattern-space, $\mathbf{m}$. The dynamics occur in pattern space and are then converted to neuron space by the stored patterns, $\{\vec{\xi}\}$. The primary constraint on the dynamical system in pattern-space is that its attractors are located on the vertices of the simplex (i.e. the points $m_{\mu} = \delta_{\mu \nu}$ are attractors, for all stored patterns $\nu$). This is generally achieved using a nonlinear function, such as sign or soft-max, that takes $m^{\nu}$ as input and limits the dynamics so that $|m^{\nu}| \leq  1$. The next section shows how one can use these ideas to think about Hopfield models as gradient-like systems.

\subsection{As traversing a landscape}\label{landscape}

\begin{figure}[ht]
\includegraphics[width=\textwidth]{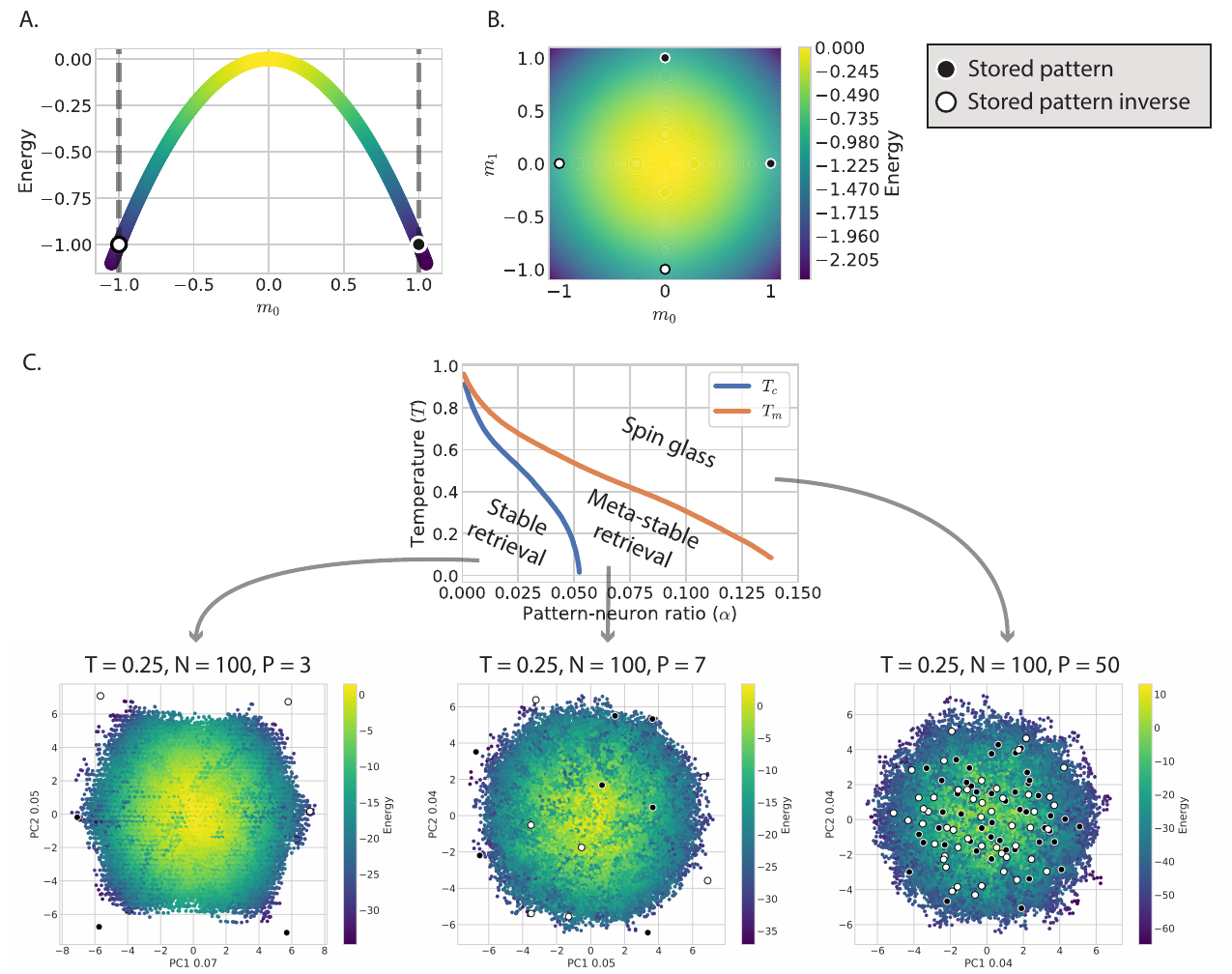}
\caption{Energy landscape of the classic Hopfield model. The energy is an inverted parabola in the space of the order parameters, $\{ m_{\mu} \}$, as in Equation \ref{eqn:energy}. The stored patterns are shown as black circles with white outlines, and the negatives of the stored patterns are shown as white circles with black outlines. A. The energy as a function of just one order parameter, $m_0$, with $m_{\mu \neq 0} =0$. The stored pattern $\xi_{\mu =0}$ and its negative $(-\xi_{\mu=0})$ are located at $m_0 = 1$ and $m_0 = -1$ respectively. These points act as attractor states because Hopfield dynamics descend the energy landscape but are restricted to $|m_0| \leq 1$ (this constraint is shown with a dashed line). B. The energy landscape in two dimensions. C. (Top) The phase diagram of the classic Hopfield model (adapted with permission from Ref. \citet{amit1985storing}. Copyrighted by the American Physical Society). The x-axis is the ratio of patterns to neurons, $\alpha = P/N$. The y-axis is the temperature of the Glauber dynamics. Above the $T_m$ curve, the model acts as a spin glass with many local minima. Below this curve, the stored patterns are meta-stable states. Below the $T_c$ curve, the stored patterns are stable states and act as attractors. This is the region of stable retrieval. (Bottom) Three PCA plots in the three phases with varying $\alpha$, at $T = 0.25$. Each plot shows many trajectories of Hopfield models, with each trajectory starting at a different random state. Each state is colored by the energy. (Left) Hopfield dynamics in the stable retrieval phase. The trajectories show an inverted parabola, as in B, with the stored patterns (and their negatives) at the global minima, or outer edges, of the parabola. (Middle) Hopfield dynamics in the meta-stable retrieval phase. The trajectories still create a landscape that looks similar to B, but now some stored patterns located in the inner part of the parabola rather than the global minima. (Right) Hopfield dynamics in the spin glass phase. The energy landscape has many local minima, and stored patterns aren't located at the global minima. The Hopfield model no longer retrieves the stored patterns.}
\label{fig:classic energy}
\end{figure}

Hopfield models are famously energy-based models: they generally have an associated Lyapunov or energy function that decreases as time passes. In other words, Hopfield dynamics descend an energy landscape and find minima corresponding to stored patterns. As previously mentioned, the energy for the classic Hopfield model with generalized order parameters is
\begin{equation}
    E^{quad} = - \frac{1}{2N} \sum_{i,\mu,\nu,j} x_i \xi_{\mu i} g^{\mu \nu} \xi_{\nu j} x_j = - \frac{N}{2} \sum_{\mu} m_{\mu} m^{\mu}.
 \label{eqn:energy}
\end{equation} 
 This energy has the same form as an Ising model with all-to-all pairwise interactions, a model that lives in ``infinite dimensions''. This makes it hard to visualize this energy landscape for a generic choice of couplings $J_{ij}$. However, Hopfield models have a built-in dimensional reduction: the order parameters $\vec{m}$. In terms of $m_\mu$ and $m^\mu$,  $E^{quad}$ is a quadratic function. As shown in Figure \ref{fig:classic energy}A and B, it is simply an inverted $P$-dimensional parabola centered at $\vec{m} = 0$. The dynamics of the Hopfield model cause the system to move down this parabola.  However, the system doesn't descend indefinitely because it is restricted by the nonlinearity in the dynamics. The trajectories are limited to those allowed by the sign, soft-max, or other nonlinear function. One of the restrictions imposed by the nonlinearity is a bound on $\vec{m}$-space, such that $|m^{\mu}| \leq 1$.

\begin{figure}[ht]
\includegraphics[width=\textwidth]{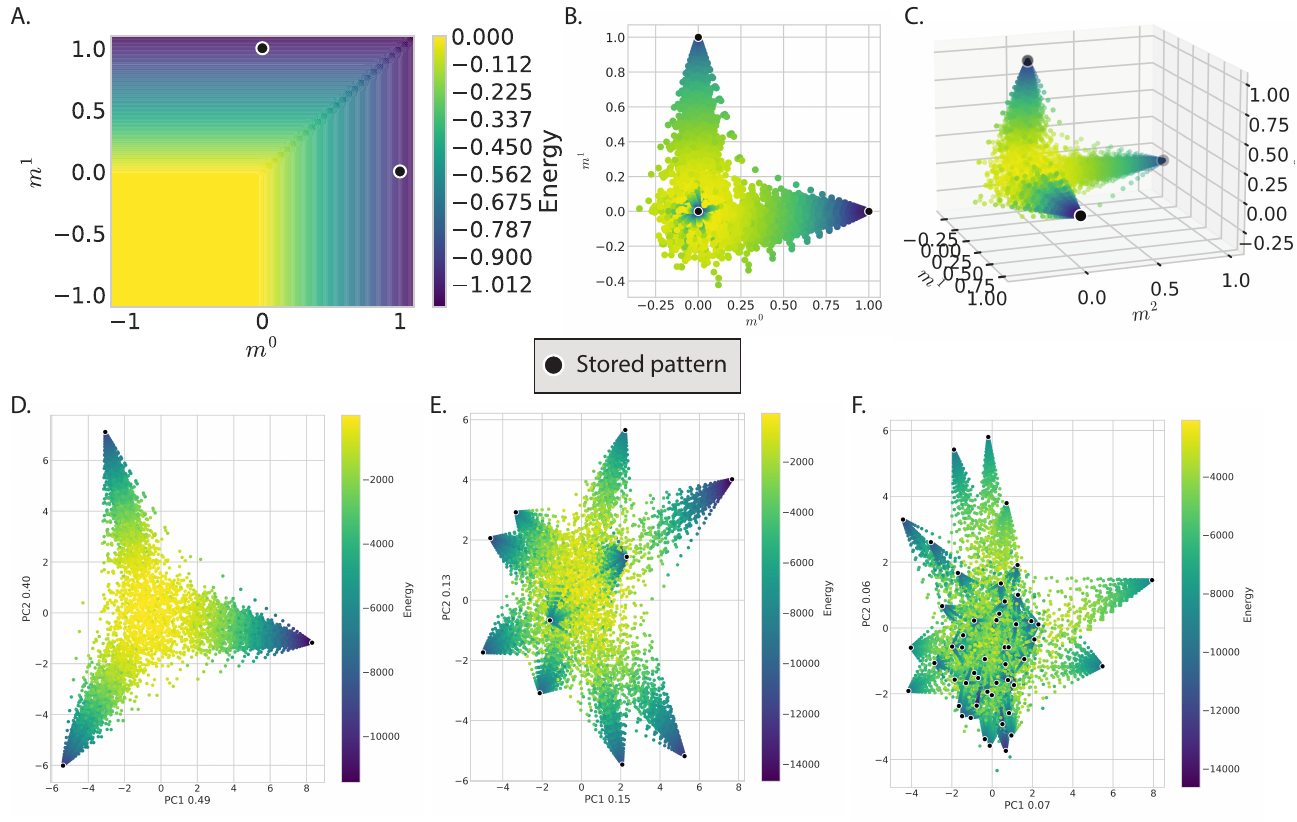}
\caption{Energy landscape of the exponential (softmax) Hopfield model. A. The log-sum-exp energy of Equation \ref{eqn:softmax energy} plotted in two dimensions as a function of order parameters $\{m^{\mu}\}$. Each stored pattern has its own basin of attraction, which are separated by energy barriers. The softmax limits the dynamics such that $\sum_{\mu} m^{\mu} \leq 1$, so that although the stored patterns aren't at the minima of the energy function, they are at the lowest energies accessible to the system. B-F: Scatter plots of trajectories colored by the classic, quadratic energy in Equation \ref{eqn:energy} show that the softmax Hopfield model also descends an inverted parabola, although the dynamics restrict the trajectories differently than in the classic Hopfield model. For each scatter plot, the continuous exponential Hopfield model is initialized at many random states and then evolved. In each system, the number of neurons is $N=100$ and the temperatures are set such that the system is well within the regime of stable pattern retrieval ($\beta = 100 \log P$). B-D. Dynamics for 3 stored patterns. B-C. Trajectories as a function of the order parameters $\{ m^{\mu}\}$ in two and three dimensions. The order parameters provide a natural dimensionality reduction of the system, with the stored patterns forming a $P-1$ simplex in this space. The stored patterns are the vertices of this simplex. D-F. PCA allows visualization of the $P-1$ simplex in two dimensions. In each case, the trajectories are restricted to the simplex while moving down the inverted parabola in Equation \ref{eqn:energy} and Figure \ref{fig:classic energy}. D. Trajectories for $P=3$. E. Trajectories for $P=10$. F. Trajectories for $P=50$. Even in the case of many stored patterns, the system achieves stable retrieval.} 
\label{fig:softmax energy}
\end{figure}

To see this, one can rewrite the dynamics of both classic and modern Hopfield models in terms of $E^{quad}$ as
\begin{align*}
    x_i^{\text{classic}} (t+1) &= \text{sign} (-\frac{1}{N}\sum_{\mu} \xi_{\mu i} {\partial E^{quad} \over \partial m_{\mu}}) \\
x_i^{\text{modern}} (t+1) &= \sum_{\mu} \xi_{\mu i} \sigma^{\mu}(-\frac{\beta}{N} {\partial E^{quad} \over \partial m_{\mu}}).
\end{align*}
This shows that the update rule of both classical and modern networks perform a gradient-like descent on the parabolic landscape defined by $E^{quad}$, but rectified through different non-linear functions (a sign function for the classical model, a soft-max function for the exponential modern network). The sharper non-linearity used in the modern Hopfield model amplifies small differences in the gradients in pattern space, allowing these networks to have a dramatically higher storage capacity than classical models. Figure \ref{fig:softmax energy}B,C shows the trajectories of the exponential modern Hopfield model in $\vec{m}$-space, with each point colored by the quadratic energy $E^{quad}$. Although the nonlinearity in the dynamics changes the trajectories, in both cases, the system ultimately descends the parabolic energy landscape defined by $E^{quad}$.

While the order parameters $\vec{m}$ provide a natural, lower-dimensional set of coordinates to observe the energy landscape, the large number of patterns being stored makes it difficult to visualize the full dynamics in pattern space. To get a broad view of a system with many attractors in one plot, one can use dimensional reduction methods like principle component analysis (PCA) \cite{Hotelling1933,PCAarticle}. Figure \ref{fig:classic energy}C and Figure \ref{fig:softmax energy}D-F use PCA to visualize the energy landscape for the classic and modern models with varying numbers of stored patterns. When the Hopfield model is in the regime of stable retrieval, the quadratic energy decreases along the trajectory of the dynamics. 

Up to this point, we have restricted our discussion of the classic Hopfield model to zero temperature $(T = 0)$. In analogy to spin systems like the Ising model, one can introduce a nonzero temperature $T$ which allows spins to flip stochastically via Glauber dynamics. This involves replacing the deterministic update rule in Eq.~\ref{Eq:Hop-dyn-gen} by a temperature-dependent stochastic rule that depends on energy differences. More precisely, at each time step, we choose a random spin $i$ and update its state at the next time step with probabilities
\begin{align*}
    Pr(x_i(t+1) = +1) &= \frac{e^{E_{+1}/T}}{e^{E_{+1}/T} + e^{E_{-1}/T}} \\
    Pr(x_i(t+1) = -1) &= \frac{e^{E_{-1}/T}}{e^{E_{+1}/T} + e^{E_{-1}/T}}
\end{align*}
where $E_{\pm 1} = E(x_i = \pm 1)$ is the Hopfield energy with the $i$th spin set to $\pm 1$. The addition of temperature introduces randomness. When $T$ is small, the Hopfield model retrieves patterns in a manner similar its zero temperature behavior. When $T$ is large, the model fails to retrieve the stored patterns and instead follows random trajectories. 

The phases of the Hopfield model at varying temperatures and pattern-packing fractions $\alpha = \frac{P}{N}$ have been studied extensively \cite{amit1985storing,PhysRevA.32.1007,amit1989modeling,mezard2017mean}. The phase diagram is shown in Figure \ref{fig:classic energy}. The Hopfield model can perform accurate pattern retrieval below the $T_c$ curve. This corresponds to a parameter region with low temperatures and small pattern packing fractions $\alpha \ll 1$. In this region of the phase diagram, the stored patterns are global minima of the energy landscape. This can be seen in the first PCA plot at the bottom of Figure \ref{fig:classic energy}C, which looks very similar to the quadratic zero-temperature landscape shown in Figure \ref{fig:classic energy}B.

 When the number of stored patterns increases to a point between the $T_c$ and $T_m$ curves, the model is in the meta-stable retrieval phase. Some, if not all, stored patterns are meta-stable states: they are local minima but no longer global minima. Even if a stored pattern is retrieved, the system may leave that state due to fluctuations. The energy landscape for the meta-stable phase is shown in the middle PCA plot of Figure \ref{fig:classic energy}C. While the energy still resembles a quadratic potential, some stored patterns are in the inner parts of the potential, at higher energy states than other points along the trajectories. Finally, the Hopfield model's pattern retrieval ability is completely destroyed in the spin glass phase, where the number of patterns exceeds the storage capacity or the temperature is too high, or both. In this phase, stored patterns aren't minima at all. In the right-most PCA plot of  Figure \ref{fig:classic energy}C, the energy landscape looks less quadratic and more irregular, and the stored patterns are scattered throughout the two principal components.

\subsection{Other perspectives}\label{other}
The literature on Hopfield models is vast and this review is far from comprehensive. In this section, we present a small sampling of other ways to understand these models. Much of the literature on modern Hopfield networks is motivated by an interest in interpretable machine learning. This is because Hopfield models are a case of neural networks that can be studied with analytical methods, and they can also be re-interpreted as Restricted Boltzmann Machines \cite{Smolensky1986,Barra2018,Mehta2019,Agliari2020}. Many tools from the statistical physics of spin glasses, such as the cavity and replica methods, have been applied to understand the behavior of Hopfield networks \cite{mezard2017mean,Agliari2023}. Additionally, there is a large body of work dedicated to finding the storage capacity and scaling of spurious states of Hopfield models with different interactions, interaction functions, learning rules, and so on \cite{Gardner1987,krotov2016dense,demircigil2017model,ramsauer2020hopfield,Agliari2023,https://doi.org/10.48550/arxiv.2304.14964, negri2023storage}.  

Before proceeding to biophysical applications we discuss one final useful perspective on these models
inspired by the bipartite structure of Restricted Boltzmann Machines and the biology of neural connections \cite{krotov2020large}. Instead of formulating Hopfield models in terms of all-to-all connected neurons $x_i$ as in previous sections, this construction involves dividing the neurons into two groups, visible neurons, $v_i$, and hidden neurons, $h_{\mu}$. Connections are allowed  between hidden and visible neurons, but no connections are allowed between neurons in the same group. In this construction, correlations between visible neurons are induced entirely by hidden units. 

The dynamics of these networks can be written as:
\begin{align*}
    \tau_f \frac{d v_i}{dt} &= \sum_{\mu}^{N_h} \xi_{\mu i} f_{\mu} - v_i + I_i \\
    f_{\mu} &= \frac{\partial L_h}{\partial h_{\mu}} \\
    \tau_h \frac{d h_{\mu}}{dt} &= \sum_{i}^{N_f} \xi_{\mu i} g_i - h_{\mu} \\
    g_i &= \frac{\partial L_v}{\partial v_i}
\end{align*}
where $\tau_f, \tau_h$ are constants that set the time scale for dynamics, $I_i$ is the input current to the visible neurons (which we set to $0$ for the Hopfield models described above), and $f_{\mu}$ and $g_i$ are nonlinear functions that describe the outputs of hidden neuron $\mu$ and visible neuron $g_i$, respectively, and are defined in terms of derivatives of ``Lagrangian functions'', denoted $L_h$ and $L_v$ for hidden and visible neurons, respectively. It is possible to show these dynamics admit a Lyapunov function of the form
\begin{align*}
    E &= [\sum_i (v_i - I_i) g_i - L_v] + [\sum_{\mu} h_{\mu} f_{\mu} - L_h] - \sum_{\mu, i} f_{\mu} \xi_{\mu i} g_i.
\end{align*}
Different choices of $L_h$ and $L_v$ give rise to different forms of Hopfield dynamics, including the classic and modern exponential models. For example, these dynamics reduce to those of modern exponential model when the Lagrangians are chosen to be
\begin{align*}
    L_h &= \log (\sum_\mu e^{h_{\mu}}) \\
    L_v &= \frac{1}{2} \sum_i v_i ^2
\end{align*}
with the additional assumption that the hidden neuron dynamics are fast ($\tau_h = 0$). In this limit, $\frac{d h_{\mu}}{dt} = 0$ and the hidden neurons  become equivalent to the order parameters because  $h_{\mu} = \sum_{i} \xi_{\mu i} g_i$.

\section{Hopfield models as lens for understanding biological function}\label{biophysics}
\begin{figure}[t]
\includegraphics[width=5in]{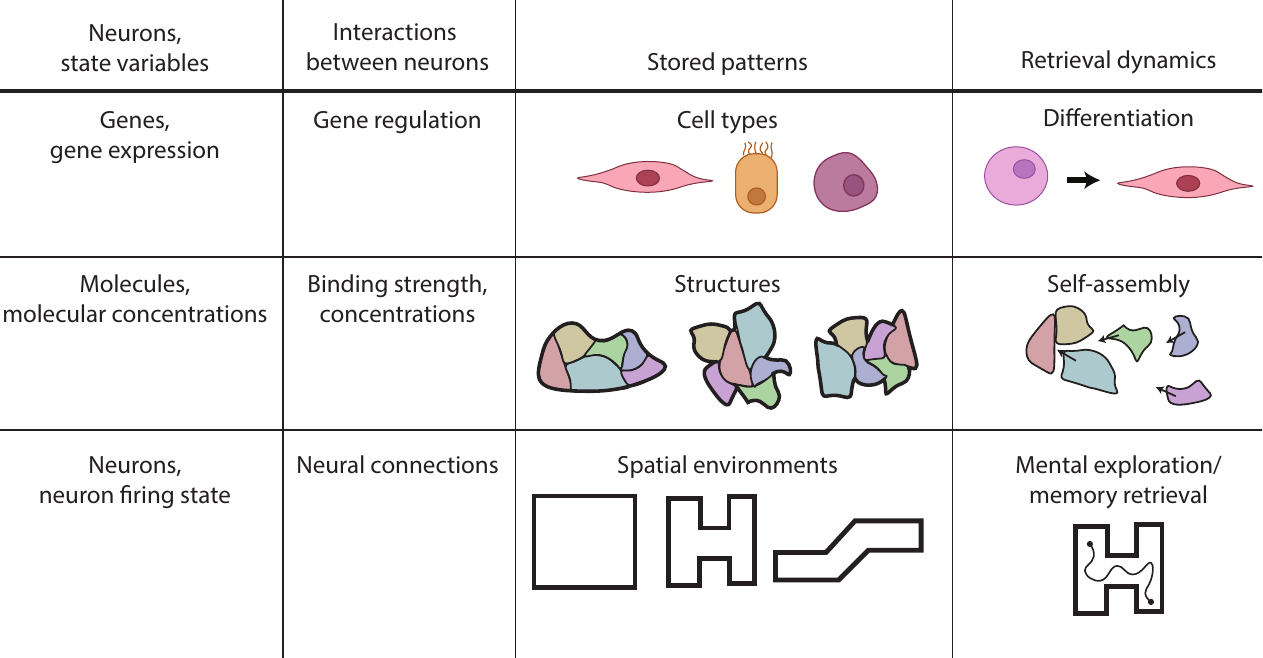}
\caption{Hopfield networks can model cell differentiation, molecular self-assembly, and spatial memory. For each of these biological contexts, the analogy to Hopfield neurons, interactions, stored patterns, and retrieval dynamics are shown. In differentiation, stem cells specify into cell types through changes in gene expression. In self-assembly, different concentrations of promiscuous molecules lead to the assembly of different stored structures. In spatial memory retrieval, neurons known as place cells are responsible for retrieving mental maps of known environments.}
\label{fig:illustrations}
\end{figure}

In this section, we discuss examples of biological systems in which Hopfield networks or associative memory have been used to model complex biophysical phenomena. We focus on three applications where this approach has been especially fruitful: (i) the epigenetics of cellular identity and cell fate transitions, (ii) self-assembly, and (iii) neural representations in the context of spatial navigation (see Figure \ref{fig:illustrations}).

\subsection{Cell fate transitions and Epigenetics}
Many animals have organs comprised of highly-specialized cells with distinct functions (e.g. lung secretory cells, red blood cells, etc.). These cell types arise during the process of development via the dynamic process of differentiation. The stability and robustness of a cell's identity once it has been acquired suggests that stable cell fates correspond to dynamical attractors of the genetic and epigenetic regulatory networks governing cellular differentiation. This idea was popularized by Waddington's now famous landscape metaphor in which the developmental pathways leading from embryonic stem cells to mature types were analogized with canals or valleys in a landscape \cite{waddington2014strategy}. Other recent works suggest that diseased states may also correspond to attractor states created by mutations or reached through a failure to transition between healthy states \cite{maetschke2014characterizing,taherian2017modeling,Wang2023}.  

There is a growing body of work attempting to formalize the Waddington metaphor using dynamical systems theory and  Hopfield models \cite{wang2011quantifying,huang2012molecular,ferrell2012bistability,moris2016transition,fard2016not,szedlak2017cell,guo2017hopland,rand2021geometry,raju2023geometrical,smart2023emergent,karin2024enhancernet,boukacem2024waddington}. In the process of differentiation, developmental gene regulatory networks use chemical and mechanical signals to guide the state of cells to particular cell fates. Like patterns in a Hopfield model, different cell types are ``stored" and then ``retrieved" by the cell (see Figure \ref{fig:interpretations}). This has inspired recent work on modeling cell identity using Hopfield-inspired  networks \cite{lang2014epigenetic, pusuluri2017cellular, smart2023emergent, karin2024enhancernet,boukacem2024waddington, Yampolskaya2025.05.27.656406}. In these models, the state of the cell, $\vec{x}$, encodes the global gene expression profile (or alternatively, the state of various developmental enhancers), with $x_i$ the expression of the $i$th gene/enhancer. The stored patterns $\{ \vec{\xi} \}$ correspond to the gene expression profiles of stable cell fates, with $\xi_{\mu i}$  the expression of the $i$th gene in the $\mu$th cell type. These gene expression profiles are often highly correlated, especially for cell types within the same organ, so it's important to use the generalized order parameter $m^{\mu}$ to describe the state of the system. In the developmental context, $m^{\mu}$ measures the similarity between a cell's current gene expression profile $\vec{x}$ and the gene expression profile characterizing cell fate $\mu$. 

The use of generalized order parameters $\{m^{\mu}\}$ as coordinates of cell fate space has proved extremely useful for analyzing high-dimensional gene expression data and is especially suited for making use of the vast amount of single-cell RNA-sequencing (scRNA-seq) data contained in developmental atlases \cite{lang2014epigenetic,pusuluri2017cellular, ikonomou2020vivo, herriges2023durable, yampolskaya2023sctop}.  In particular, the use of Hopfield-inspired order parameters provides a powerful and interpretable alternative to stochastic dimensional reduction techniques like t-SNE and UMAP for analyzing and visualizing gene expression data \cite{van2008visualizing,mcinnes2018umap,chari2023specious}. By calculating the generalized order parameters $m_\mu$ for cells, we can assign every cell a unique coordinate in  cell fate space. This coordinate system can be used both for visualization and classification of cells into cell fates. Because the stored memories are calculated directly from  single-cell atlases and are fixed, order parameter based methods also allow for robust inferences that are consistent across samples and time \cite{yampolskaya2023sctop}. 

Whereas the previous works focus on gene expression, recent work by \citet{owen2023design} use the lens of the Hopfield models to explore how the three-dimensional organization of chromatin allows cells to implement epigenetic memory. Within the model, chromatin regions play a role analogous to neurons, with connections between regions controlled by chromatin modifications. These couplings are learned through a ``mark together, park together'' learning rule analogous to the ``fire together, wire together'' rule in Eq.~\ref{Eq:Hop-J-gen}. Finally, the mark dynamics play a role similar to the dynamic update rule, Eq.~\ref{Eq:Hop-dyn-gen}. 

\subsection{Self-assembly}

A growing body of work has begun to reveal a deep and surprising connection between the physics of self-assembly and the dynamics of associative memory in neural networks \cite{murugan2015multifarious, zhong2017associative, chalk2024learning, evans2024pattern, braz2024liquid}. At the heart of this connection lies a simple but powerful idea: just as a Hopfield network retrieves a stored memory from a partial cue, a multicomponent molecular system can self-organize into a target structure in response to incomplete or noisy chemical inputs.  Though these works largely focus on simpler {\it in vitro} biophysical and soft-matter systems, they suggest that the uncanny ability of biological systems to reuse the same components to form different structures may also be governed by Hopfield-like design principles.

In these systems, the roles of neurons and couplings are played by physical components and their interactions. The state of the system at any time can be described by the concentration of molecular components -- for example, different species of proteins, DNA strands, chemical building blocks, or concentration of a liquid component.  Couplings between neurons are replaced by pairwise interaction strengths between components. A major new source of complication in these systems is nonspecific binding between components. This non-specific binding causes interference between target structures analogous to interference between patterns in the classic Hopfield model, ultimately limiting the capacity of these networks to form disparate structures.

\subsection{Neural Representations} Finally, we briefly note that there has been a renewed interest in Hopfield models in the context of neural representations of spatial structures in the hippocampus \cite{whittington2020tolman, whittington2021relating} (see  \cite{whittington2022build} for a recent review of these developments). In this context, the stored patterns correspond to internal maps of different locations that have been learned by the individual. Retrieval of a pattern corresponds to mentally exploring an internal map when presented with initial conditions similar to a known environment.While a detailed account of this literature is beyond the scope of this review, there are several interesting developments in the debates on cognitive maps that are likely to be of interest to the broader biophysical community. 

This work suggests that Hopfield models can generalize beyond memorizing patterns in order to learn latent representations.  The raises the question of how Hopfield-like models generalize to new memories that were not explicitly {\it a priori} specified? \cite{whittington2022build, whittington2021relating}? This question has also been recently explored from a statistical physics perspective in \cite{negri2023storage,kalaj2024random}. In many biophysical applications, we would like the system to function for a whole class of patterns, structures, memories, or cell types that are drawn from a probabilistic ensemble with shared characteristics. How can we generalize Hopfield models to account for this by learning latent representations in addition to memorizing patterns? 

The second concerns the ability to extend Hopfield models to include additional sensory inputs and control signals that shape stored memories \cite{whittington2022build}. In the context of spatial neural representations, this often involves debates about if, and how, inputs from the cortex bind together memories stored by the hippocampus. This raises the more general question how one can modify Hopfield models to account for landscapes where memories are dynamic, with attractors becoming stable and unstable depending on external cues. A similar question has recently been explored by \cite{Yampolskaya2025.05.27.656406} in the context of cell fate identity where developmental signals play a role analogous to cortex signals. Both these areas represent interesting directions for further investigation.
 
\section{Conclusion}

Hopfield models offer a striking example of how simple dynamical rules in high-dimensional systems can give rise to robust, emergent function in biology. These models provide a powerful lens for understanding phenomena ranging from cell fate decisions, to molecular self-assembly, to neural representation. In this review, we have presented a self-contained and pedagogical introduction to this rich body of work, with an emphasis on building intuition by exploring multiple perspectives through which to view these models. While the mathematical details may vary across systems, this conceptual lens reveals common biophysical principles that underlie emergent function in diverse biological contexts. More broadly, Hopfield models serve as a compelling example of how  a well-posed mathematical model can connect the properties of microscopic interactions to macroscopic behaviors. As we continue to search for organizing principles that govern collective dynamics in biology, the Hopfield model remains an essential conceptual tool for understanding emergent function in biophysical systems.% The following commented section is from the Annual Reviews LaTeX template

\section*{DISCLOSURE STATEMENT}
The authors are not aware of any affiliations, memberships, funding, or financial holdings that might be perceived as affecting the objectivity of this review. 

% Acknowledgements
\section*{ACKNOWLEDGMENTS}
We would like our collaborators, members of the Mehta and Kondev groups, Kotton Lab, and Laertis Ikonomou for many useful discussions that influence this review. This work was funded by The work was funded by grants from a  Chan Zuckerberg Initiative (CZI) Investigator Grant in Theory and Biology to PM and NIH NIGMS 1R35GM119461 to PM.

% References
%
% Margin notes within bibliography
% \section*{LITERATURE\ CITED}

% To download the appropriate bibliography style file, please see \url{https://www.annualreviews.org/page/authors/general-information}. 

% \\

% \noindent
% Please see the Style Guide document for instructions on preparing your Literature Cited.

% The citations should be numbered in alphabetical order, with titles.

\bibliography{bibliography.bib}
\bibliographystyle{plainnat}
\bibliographystyle{ar-style6.bst}

\end{document}